# An *In Situ* Measurement System for Characterizing Orbital Debris

Michael A. Tsao, Hau T. Ngo, *Member, IEEE*, Robert D. Corsaro,
and Christopher R. Anderson, *Senior Member, IEEE*

*Abstract*—This paper presents the development of an *in situ* measurement system known as the Debris Resistive Acoustic Grid Orbital Navy/NASA Sensor (DRAGONS). The DRAGONS system is designed to detect impacts caused by particles ranging from 50 $\mu$m to 1 mm at both low-earth and geostationary orbits. DRAGONS utilizes a combination of low-cost sensor technologies to facilitate accurate measurements and approximations of the size, velocity, and angle of impacting micrometeoroids and orbital debris (MMOD). Two thin layers of kapton sheets with resistive traces are used to detect the changes in resistance that are directly proportional to the impacting force caused by the fast-traveling particles. Four polyvinylidene fluoride-based sensors are positioned in the back of each kapton sheet to measure acoustic strain caused by an impact. The electronic hardware module that controls all operations employs a low-power, modular, and compact design that enables it to be installed as a low-resource load on a host satellite. Laboratory results demonstrate that in addition to having the ability to detect an impact event, the DRAGONS system can determine impact location, speed, and angle of impact with a mean error of 1.4 cm, 0.2 km/s, and 5°. The DRAGONS system could be deployed as an add-on subsystem of a payload to enable a real-time, in-depth study of the properties of MMOD.

## I. Introduction

ORBITAL debris, colloquially known as "space dust," has a long history of damaging space assets and is an increasing issue for current and future space missions [3]. Orbital debris can cause considerable damage due to its typical impact speeds of 10 km/s in low-earth orbit (LEO) [4]. Even debris the size of dust can cause serious damage when it travels at these velocities. In 1996, the gravity-gradient stabilization boom of an operational French satellite (Cerise) was cut in half by tracked debris, which left the satellite severely damaged and its performance severely compromised [5]. Although orbital debris has long been recognized as a concern for satellite and satellite instrumentation designers [6], it has only been recently recognized as a critical component of the design process for all spacecraft and spacecraft subsystems [7]. Until recently, shielding design for spacecraft was a secondary concern [6] and was nonexistent on older spacecraft, such as the space shuttle [7]. However, orbital debris threat assessment and shielding design have become a key concern, in order to ensure satellite and space instrument survivability [7]. Moreover, the increasing threat of orbital debris has caused a significant spike in costly collision avoidance maneuvers (CAMs) of satellites. From 1998 to 2010, the International Space Station (ISS) averaged only one CAM per year. From April 2011 to April 2012, the ISS was forced to execute four CAMs and would have conducted two additional maneuvers if the warnings had come sooner [8]. Each CAM requires extensive planning and fuel to be carried out and inhibits ISS experiments that require a continuous zero-gravity environment. The former NASA chief scientist for orbital debris, Nicholas Johnson, stated, "The greatest risk to space missions comes from nontrackable debris" [9]. Due to the increasing population of debris traveling at hypervelocity, the probability of particle impacts with mission-critical subsystems that leads to catastrophic cascade failure on a satellite or the ISS is rising.

The growth in orbital debris is largely due to man-made system and spaceflight miscalculations. Fig. 1 displays orbital debris tracked by the U.S. Space Surveillance Network (U.S.-SSN) from 1957 to 2013 [10]. The two highlighted spikes in debris were caused by two events: in 2009, the fully operational Iridium 33 was destroyed by the retired Russian Cosmos satellite [11], and in 2007, the Chinese performed an antisatellite (ASAT) test and destroyed their old weather satellites, the FY-1C. Both occurrences happened at roughly the same altitude (Chinese ASAT: mean altitude of 865 km; Iridium Cosmos: 792 km), polluting LEO with mass amounts of orbital debris that for the sake of ongoing and future space missions must be tracked and accounted for [1].

### A. Tracking Orbital Debris

Current orbital-debris models are generated by the U.S.-SSN and the NASA Orbital Debris Office. Orbital-

Manuscript received April 1, 2016; revised June 8, 2016; accepted July 10, 2016. This work was supported by the NASA Orbital Debris Program Office at the Johnson Space Center. This paper was presented at the IEEE International Instrumentation and Measurement Technology Conference, May 2012, the Joint Conference of 30th International Symposium on Space Technology and Science, 34th International Electric Propulsion Conference, and the 6th Nano-Satellite Symposium, December 2015. The Associate Editor coordinating the review process was Dr. Mark Yeary. *(Corresponding author: Christopher R. Anderson.)*
M. A. Tsao is with Northrop Grumman, Baltimore, MD 21224 USA (e-mail: miketsao@icloud.com).
H. T. Ngo and C. R. Anderson are with the Wireless Measurements Group, Electrical and Computer Engineering Department, United States Naval Academy, Annapolis, MD 21402 USA (e-mail: ngo@usna.edu; canderso@usna.edu).
R. D. Corsaro is with the U.S. Naval Research Laboratory, Washington, DC 20375 USA (e-mail: corsaro@me.com).
Color versions of one or more of the figures in this paper are available online at http://ieeexplore.ieee.org.
Digital Object Identifier 10.1109/TIM.2016.2599458

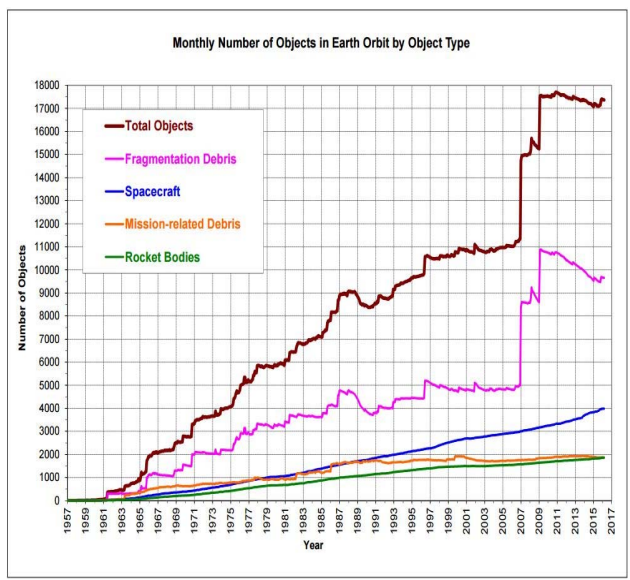

Fig. 1. Monthly number of cataloged objects in earth orbit by object type, 1957–2016, from [12].

debris measurements are accomplished by ground-based and space-based observations. Ground-based measurements consist of radar and optical systems. Typically, radar measurements have been used to measure medium-sized space debris (approximate diameter of 5 mm–30 cm) in LEO, whereas optical measurements have been used for large space debris (approximate diameter of <1 m) in high-earth orbit (HEO) [13]. Space-based *in situ* impact sampling has been the only effective means of measuring small debris flux (approximate diameter of <1mm). Space-based observations attempt to characterize the submillimeter micrometeoroids and orbital-debris (MMOD) flux through the analysis of spacecraft surfaces returned from space (for cost reasons, surfaces are retrieved only from LEO). However, this method is inefficient due to the time and cost of retrieving a spacecraft surface. Roughly speaking, there are about 21 000 particles greater than 10 cm in diameter, 500 000 medium particles greater than 1 cm in diameter, and an estimated tens of millions of particles less than 1 cm in diameter [3], [13], [14]. MMOD is both natural and man-made, with the man-made components consisting of paint flakes, dust, solid rocket motor slag, and satellite fragmentation [3]. Traveling at hypervelocities of greater than 5 km/s, these particles can cause catastrophic damage to solar panels, satellite or space vehicle surfaces, and can even cripple an entire subsystem. The potential for damage from untracked MMOD is unpredictable, and there is thus an increasing need for an efficient and more comprehensive method of detecting and quantifying the MMOD environment in both LEO and HEO.

*B. MMOD Threat Assessment*

In the U.S., the NASA Orbital Debris Office assesses the level of MMOD threat through a variety of means. In 1984, NASA launched a bus-sized spacecraft called the Long Duration Exposure Facility designed to characterize the effects of MMOD on various spacecraft materials [15]. After 5.7 years in LEO and over 20 000 documented impacts, it was determined that MMOD impacts significantly degrade the performance of exposed spacecraft materials and, in some cases, destroy a satellite's ability to perform or complete its mission. Allbrooks and Atkinson [15] state that "these types of impacts on deployment mechanisms, for example, could prevent them from operating and could prevent the retraction or the initial deployment of things, such as solar panels or communications antennae." In addition, MMOD have damaged critical surfaces and satellite subsystems, such as shuttle windows, Hubble Space Telescope high-gain antenna, Small Expendable Deployer System-2 tether, and other exposed shuttle surfaces [13]. Due to the vast number of MMOD impacts to which a surface can be subjected, there have been many satellite anomalies that many researchers believe are caused by MMOD, as detailed in [16]. Because there is a lack of concrete knowledge regarding the population and distribution of MMOD, the threat of MMOD to space missions requires a method of measuring and characterizing these particles.

The result of this highly detailed MMOD threat assessment was the development of the NASA Orbital Debris Engineering Model (ORDEM), which is used to statistically characterize the risk of orbital debris to spacecraft designers and operators [17]. The population and flux of MMOD in this model, however, was principally updated through the examination of returned spacecraft surfaces via the U.S. Space Shuttle. Because the Space Shuttle was retired from service in 2011, a critical gap in the ORDEM knowledge of the population, size, distribution, and flux of MMOD has developed, leading to an increasing risk of the catastrophic failure of space assets. In addition, because the population of satellites continues to grow, there is an increasing need for real-time *in situ* situational awareness of the MMOD environment, in order to achieve more accurate risk assessment and collision avoidance.

*C. Previous and Current Efforts*

Space-based instrumentation has long been of interest to scientists and engineers, particularly when considering satellite-based instrumentation. In the early days of space exploration, Flowerday [18] was concerned with characterizing the atomic composition of the upper atmospheric and low-earth orbital environments. This instrument was intended to characterize the ionic structure of the upper atmosphere to better understand and predict long-distance high-frequency radio wave propagation [18]. Keppler *et al.* [19] developed an instrument for characterizing charged particle flux in LEO, with an emphasis on understanding the threat to electronic components from the total radiation does and single-event upsets (SEUs). They were interested in the measurements of particle mass, velocity, and energy—similar to the MMOD measurements performed by Debris Resistive Acoustic Grid Orbital Navy/NASA Sensor (DRAGONS)—and use a similar multilayer sensor design and interleaved sampling system [19]. Researchers are also extremely interested in instruments for measuring and monitoring space structures, such as the ISS. For example, Rice *et al.* [20] developed an instrument for precision measurements of the microgravity environment on the ISS.

Their system was designed to correlate microgravity changes to both experiment anomalies and vehicle disturbances (such as impacts by MMOD), and illustrates the requirements of autonomous operation, performance verification, and on-orbit validation inherent to any space-based instrumentation [20].

The vast majority of efforts to characterize MMOD have been *in situ* measurements. Researchers have attempted to accomplish such measurements using a variety of impact-detection methods: acoustic sensors, impact ionization sensors, optical sensors, conductive traces, and aerogel. One of the earliest projects to measure MMOD was the Debris In-orbit Evaluator (DEBIE), which was launched in 2001 on-board the European Space Agency Project for On-Board Autonomy. DEBIE characterized MMOD using impact ionization sensors and acoustic-strain sensors [21]. The impact ionization sensors determined MMOD mass and velocity, whereas the acoustic sensors determined the momentum. However, the system's plasma sensors performance suffered from false triggering [22].

Between 2003 and 2005, research was carried out on a modern acoustic sensor for MMOD impact detection called the Particle Impact Noise Detection and Ranging on Autonomous Platform (PINDROP) [23]. The idea was to combine acoustic sensors with a collection tray to characterize and capture MMOD. The acoustic sensors were attached to the collection tray to enable the measurement of the composition and physical characteristics of captured particles. After extensive testing, it was determined that, for this application, polyvinylidene fluoride (PVDF) was the best available sensor material for measuring MMOD. PINDROP was selected to be a component on the Large Area Debris Collector, an experimental module that was intended to collect and analyze MMOD on the ISS. However, this project was terminated due to budget constraints in early 2007 [24].

In 2014, the Naval Research Laboratory investigated developing an optical orbital-debris spotter. The measurement concept behind the debris spotter is to create a light sheet using a low-power laser and a conic mirror. When particles intersect the light sheet, the particles reflect, transmit, and/or absorb the light [25]. This disturbance in the light sheet is detected by a camera and is expected to determine MMOD size and shape, as well as the distance of MMOD from the sensor. It is expected that the optical orbital-debris spotter concept will be able to measure particles with a diameter as small as 0.01 cm.

Currently, under research at Kyushu University is an MMOD detector called *In situ* Debris Environmental Awareness (IDEA) [26]. The objective of IDEA is to deploy a group of small piggyback satellites equipped with dust sensors to monitor congested orbital-debris environments. IDEA uses a "dust sensor" that consists of a multitude of thin, conductive traces that are 70 $\mu$m wide with 30 $\mu$m spacing in a linear array on a thin substrate. IDEA detects particle size by determining the number of broken conductive strips caused by an impact.

The ISS experiments conducted by the National Space Development Agency of Japan called microparticle capturer (MPAC) and Space Environment Exposure Device collected data from 2001 to 2004 [27]. MPAC was composed of aerogel and polymide foam for particle capture, and it had an aluminum back plate for crater counting. Three units were installed on the ISS in 2001, and they were retrieved individually at approximately one year intervals, in 2002, 2003, and 2004. Although the experiment provided valuable data, it lacked *in situ* data-gathering capability, because the sensors were required to be returned to earth for manual inspection.

The DRAGONS is a current MMOD measurement project that builds upon past research by combining three different technologies to maximize the information extracted from each detected impact. DRAGONS uses acoustic-strain sensors, a resistive grid, and a dual-layer system. The combination of these three technologies enables the measurement of the size, speed, and angle of arrival (AoA) of impacting MMOD. The goal of the DRAGONS project is to create a low-power/ low-resource, inexpensive independent spacecraft instrument to fill the measurement gap for particles ranging from 50 $\mu$m to 1 mm at both LEO and Geosynchronous Orbit. DRAGONS would interface with the host satellite via a system such as the Space Environment Testbed (SET) [28]. The SET—or similar bus—provides power, telemetry, housekeeping, and a host interface for command and control to multiple satellite experiments or instruments, thus providing a common interface for a multiexperiment payload.

## II. Instrument Design

Fig. 2(a) shows the DRAGONS sensor and structure. The sensor consists of a mechanical structure that has two thin kapton sheets stretched across the structure to be flush, flat, and unwrinkled, but not under tension, with a 15-cm vertical separation between the sheets. The outer sheet is a resistive grid, which is divided into quadrants, on each of which is printed numerous parallel resistive traces. On the back of each sheet, four PVDF acoustic-strain sensors are positioned in the corners, equidistant from the center of the film. The measurement process begins when an MMOD impact penetrates the two films. The impact causes a longitudinal propagating wave to be transmitted across each sheet; this lateral compression and elongation (transverse acoustic wave) are measured by the acoustic-strain sensors. The hole created by the particle impact on the resistive sheet severs some of the traces, causing a change in resistance directly proportional to the size of the hole.

The DRAGONS instrument's electronics consists of three subsystems: the acoustic subsystem (ACS), the resistive-grid subsystem (RGS), and the control and data storage subsystem (CDSS). The ACS serves as the primary measurement system; it records the longitudinal propagating signal output from the eight acoustic-strain sensors. Measuring the arrival time and magnitude of each signal enables postprocessing analysis to calculate location, AoA, and relative speed of an impacting particle. The RGS functions as the complementary measurement system to the ACS, by determining the approximate size and location of an impact as well as confirming its occurrence via a measured change in resistance. The CDSS stores the measurement data of the ACS and RGS and interfaces with the host satellite to enable the ground system to download and process the data gathered. A detailed overview of the DRAGONS system operation can be found in the Appendix.

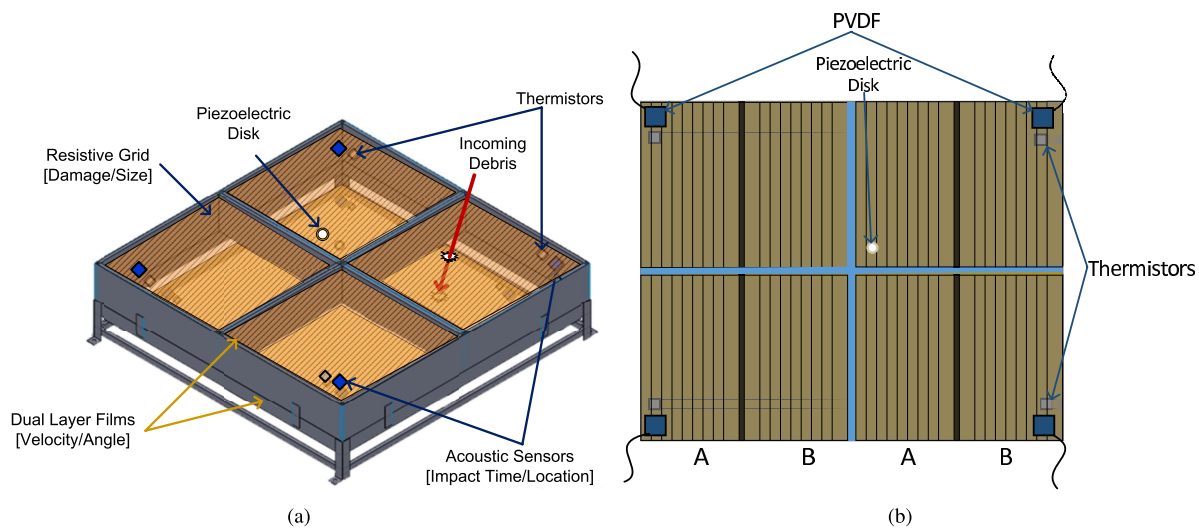

Fig. 2. DRAGONS instrument sensor. (a) Conceptual illustration of the DRAGONS instrument showing the dual-layer mechanical structure with top-layer resistive-grid sensor, the placement of acoustic sensors, the placement of the thermistors, and nominal placement of the piezoelectric disk actuators. (b) Illustration of the resistive-grid sensor, placement of the PVDF acoustic sensors, thermistors, and piezoelectric disk actuator.

### A. DRAGONS Resistive/Acoustic Sensor

The DRAGONS resistive sensor is shown in Fig. 2(b). The resistive sensor consists of four quadrants, each of which have 1640 parallel resistive traces. Each quadrant is split into two resistive grids: grids A and B (each having 820 traces). The traces are made from nickel chromium aluminum silicon (NCAS) printed onto Kapton Film. NCAS was selected, because it has an extremely small resistance change over temperature, with a temperature coefficient of resistance of 20 ppm/°C. This highly stable resistance over temperature is critical, because DRAGONS will operate in an environment, where it will experience temperature variations from $-60$ °C to $+95$ °C. The NCAS traces have a resistivity of 100 $\Omega$ per square, and they are configured to be 75 $\mu$m wide with a 75 $\mu$m gap between traces–the smallest configuration possible with NCAS and current manufacturing techniques. The RGS detects particle impacts by monitoring the changes in resistance when traces are broken. Based on the number of resistive traces broken on one of the four resistive quadrants, the RGS is able to determine the size and general location of MMOD. To track and calibrate out any temperature fluctuations, thermistors were placed in the four corners of each of the kapton sheets.

The acoustic-strain sensors on each sheet produce a voltage proportional to the magnitude of the acoustic waves on the sheet, which allows DRAGONS to calculate a number of particle characteristics. The magnitude of the acoustic waveform allows analysis to estimate the size of the impacting particle. In order to calculate impact location and AoA, a modified multilateration time-of-arrival (ToA) process is used. The multilateration process determines the impact location using the difference in distance to each sensor. The impact distance from each sensor is determined by using the ToA to each sensor and the known wave speed on the material. Only three sensors are needed for the multilateration process, and the fourth sensor is used to increase accuracy. In addition, piezoelectric disk actuators are placed on both katpon sheet layers to calibrate the ACS. The piezoelectric disk can be activated to generate acoustics vibrations along the film to provide a running relative calibration of the acoustic-strain sensors to ensure that they have not been compromised. With a fixed distance to each sensor, the actuator can be used to monitor any changes of wave speed on the film. More information on the use of the piezoelectric disk for calibration purposes can be found in Section IV-B.

### B. Electronics Subsystem

Fig. 3 displays the block diagram for the three subsystems. The ACS is responsible for processing and temporarily storing the acoustic signals. The processing chain begins with the outputs of the eight PVDF acoustic-strain transducers. The eight acoustic signals are high-pass filtered with a cutoff frequency of 11 kHz to eliminate ambient noise. Mechanical vibrations generally occupy the lower frequencies and provide no signals of interest for impact detection. After filtering, each signal is amplified to occupy the full-scale input range of the analog-to-digital converter (ADC). Each acoustic signal is time multiplexed through an 8-to-1 analog multiplexer (MUX). The eight channels of the MUX are switched at a rate of 4 MHz, which results in a sampling frequency of 500 kHz per channel. The typical propagation delay of the MUX is 95 ns, which is well within the switching, settling, and hold time of the MUX-ADC combination at the 500 kHz sampling frequency. It was empirically determined that 500 kHz was the minimum necessary to ensure fidelity of the acoustic data necessary to characterize the impacting particles. The analog output of the MUX is then sampled and digitized by a single 12-b ADC.

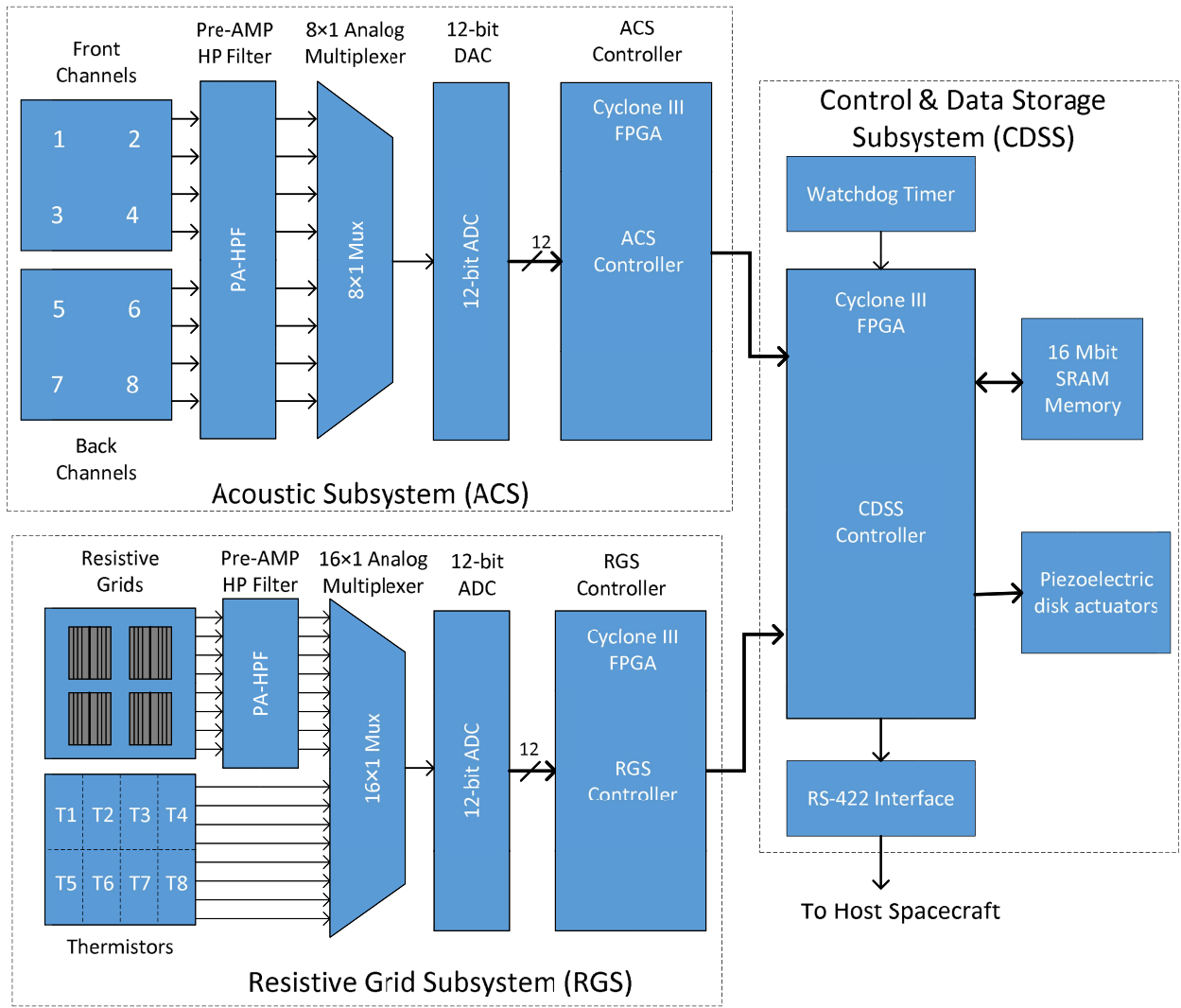

Fig. 3. Block diagram of the DRAGONS electronics system, illustrating the ACS, RGS, and CDSS.

The acoustic data are constantly sampled and stored in a first-in first-out (FIFO) buffer within the field-programmable gate array (FPGA). When a digitized acoustic sample crosses a specified impact-detection threshold, the ACS saves preimpact data already stored in the FIFO and records the postimpact data to be transferred to the CDSS for storage.

The RGS is responsible for measuring the resistive grids, measuring the thermistors, and temporarily storing the data. The RGS measures three resistances for each of the quadrants: the individual resistances of each grid (A and B) and a differential measurement between the two grids. The changes in resistance measurements of grids A and B are used to confirm the impact and approximate location of a particle. The differential measurement of the two grids is used for a high-resolution measurement of the actual number of resistive traces that were broken by the impact. Given the NCAS resistivity of 100 $\Omega$ per square, each 75 $\mu$m × 25 cm trace has a nominal resistivity of 330 k$\Omega$, and each 820-line resistive-grid starts with a nominal resistance of 400 $\Omega$. To measure the resistance, a constant-current source injects 3 mA into each grid, which produces a voltage proportional to resistance. Low-noise precision operational amplifier circuits are then used to measure the sum $(A + B)$ and difference $(A - B)$ of voltages produced by each grid; by measuring and recording the sum and the difference, the system can calculate individual resistances of grids A and B, obtaining the three required data points with only two measurements.

The RGS board has 16 input channels: eight for measuring the resistive grids and eight for measuring the thermistors. The signals are an input to a 16-to-1 MUX, the output of which is sampled by the ADC in a sequential step-through process. This process averages 4096 samples of a single channel, stores the average, switches to another channel, and then repeats this cycle through all 16 channels. Because each line break results in a resistance change of about 0.7 $\Omega$, a precise measurement is required in order to detect a single line break.[1] Therefore, a single measurement is an averaged value; this mitigates the presence of thermal and instrument noise. Similar to the ACS, the RGS data are stored within an FIFO in the FPGA before being transferred to the CDSS.

---

[1]Note that an impacting particle with the diameter of less than 50 $\mu$m will cause a partial line break and result in a change in resistance of less than 0.1 $\Omega$. This change is very near the minimum resolution of the ADC, and would thus not be recorded as a "line break" by DRAGONS.

The CDSS is the nucleus of DRAGONS. The CDSS is responsible for data storage, power regulation, in-flight calibration, health and status checks, time and date tracking, and communication with the host satellite. It stores the measurement data from the ACS and RGS in a static random access memory (SRAM) chip. The CDSS controls the periodic calibration measurements required of the ACS and RGS. For the ACS, the CDSS controls two piezoelectric actuators that generate a 20-kHz acoustic pulse on the two films. For the RGS, the CDSS records a resistance and temperature measurement from the RGS to monitor resistive changes due to the environment during orbit. In order to track the orbital position of events, the CDSS tracks the date and time with a real-time clock synchronized once per second to the host satellite time. A watchdog timer also periodically generates a signal to check if the system is still functioning. Finally, the CDSS functions as the link between the DRAGONS system and the host spacecraft. The CDSS sends and receives commands to and from the host satellite when actions such as a data upload is necessary.

FPGA devices are selected to implement the digital controllers on the ACS, RGS, and CDSS subsystems. FPGA devices with embedded memory blocks, registers, high-speed memory and storage interfaces, and large numbers of input/output pins provide suitable platforms to rapidly realize cost-effective on-board controllers. Since these FGPA devices can be configurable and reconfigurable, they offer great flexibility to adapt to changing requirements of the system to support future updates and upgrades. FPGA devices have been widely used to implement controllers in embedded systems to record data and monitor critical components in various industrial and space-based applications [29]–[32]. The configurability and availability of interconnected logic cells in an FPGA are particularly well suited for computationally and data intensive applications [33], [34]. Since most space-based hardware systems have limited on-board resources reserved for data processing, FPGA devices provide attractive solutions for low cost, dedicated on-board capability for real-time computations [35]–[38]. With an embedded FPGA device on-board, each of the ACS, RGS, and CDSS subsystem provides significant processing capability that can be utilized to implement various signal analysis algorithms if needed.

As with any electronic systems deployed in space, FPGA devices are susceptible to the harsh environment that may cause radiation-induced errors known as SEUs. An SEU error in a microelectronic system will cause a bit flip or a bit constant (stuck at ground or at a high voltage level). This error could cause significant degradation to the overall performance if it occurs in one of the critical components of the electronic system (e.g., the digital controller). As the use of FPGA devices in space-based systems increases, the need to improve the reliability of these systems is also increasing. There are many different methods to mitigate the effects of SEU errors in FPGA-based designs. One of the methods is to use the on-chip resources that are dedicated for SEU mitigation technologies. For example, FPGA devices from Cyclone II family that are used in this paper include on-chip automatic cyclical redundancy check (CRC) code to detect an SEU error in the configuration bits. If an error is detected, a reconfiguration of the FPGA chip can be triggered automatically [39]. Another method is to implement a well-known Secure Hash Algorithm to detect and identify an SEU error [40]. The use of a classic technique in fault-tolerant design known as triple modular redundancy (TMR) is another common approach [41]. With the TMR technique, the designs of critical components are replicated three times. The three replicas are used concurrently in conjunction with a majority voting scheme to eliminate an error. This technique has been proved to improve the reliability of fault-tolerant systems [41]. In this paper, we use a combination of two approaches to mitigate SEU errors: on-chip CRC and TMR technique.

TABLE I
RGS MEASURED PERFORMANCE RESULTS

| Parameter | Value |
|---|---|
| Difference Linearity | $R^2 = 0.9995$ |
| Sum Linearity | $R^2 = 0.9999$ |
| Temperature Dependence 0°C–70°C | $\pm 18\ m\Omega$ |

## III. SYSTEM VALIDATION

A comprehensive system performance testing was conducted on a flight prototype of the DRAGONS electronics system to determine if the system could accurately and consistently measure particles from 50 to 1000 $\mu$m in diameter. Initial isolated testing was conducted on each subsystem using calibrated inputs. When this was completed, the subsystems were integrated, and the DRAGONS performance as a whole was validated using a low-velocity air gun to simulate a particle impact.

### A. RGS Validation

Preliminary validation testing was completed on the prior version of the electronics system and published in [1]; a more comprehensive evaluation of the resistive grid and DRAGONS structure was published in [2]. For the flight-prototype unit, the RGS was subjected to the same tests as detailed in [1]: linearity, sensitivity, and temperature interdependency; results are summarized in Table I. The linearity test was to confirm the linear relationship between the input voltage of each analog input channel of the RGS and the digital output produced by the ADC. This test ensured that the resistance could be accurately and precisely calculated using the digital output produced from the ADC. The sensitivity test was conducted to validate the system ability to detect a 50-$\mu$m particle, corresponding to a single 0.7 $\Omega$ change in resistance. For these two tests, the RGS was connected to a simulated resistive grid composed of a series of various-valued precision 1% tolerance resistors connected in parallel. Resistance changes could be emulated by removing one or more resistors from the parallel combination. For the test, a variety of different resistance values were chosen in the expected input range of the RGS, $[-100\ \Omega, 100\ \Omega]$ for the difference circuit and $[800\ \Omega, 1200\ \Omega]$ for the summing circuit; actual resistance values were measured using a precision 6.5 digit benchtop multimeter. A linear regression was then performed on the measured data; the results are shown in Fig. 4. Fig. 4 shows that: 1) both the sum and difference measurements are extremely linear over the entire range of expected input resistance values and ADC output codes and 2) the RGS maintains linearity even when detecting a 0.5 $\Omega$ change in resistance, as evidenced in Fig. 4 (inset).

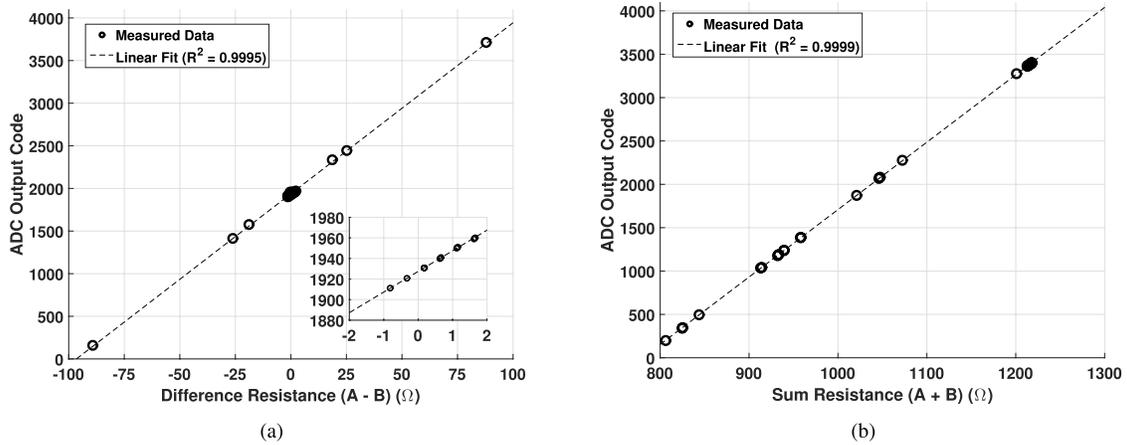

Fig. 4. Results of linearity and sensitivity testing of the DRAGONS RGS subsystem. (a) Performance of the difference circuit demonstrating linearity to measure $(A - B)$ across the full range of expected inputs of $[-100\ \Omega, 100\ \Omega]$, as well as the demonstrating sensitivity to measure a single line break (inset). (b) Performance of the summing circuit demonstrating linearity to measure $(A + B)$ across the full range of expected inputs of $[800\ \Omega, 1200\ \Omega]$.

Temperature variability was another concern, because temperature-dependent changes in the electronics could result in an erroneous recording of a resistance value. Although changes could be introduced by any element in the electronics system, of particular interest was the temperature stability of the 3-mA constant-current source, which used a temperature compensation circuit to account for the positive temperature coefficient of the current regulator. To test the temperature variability, the RGS electronics were placed inside a temperature chamber. A precision 500-$\Omega$ resistor was connected to the RGS to emulate a resistive grid and placed outside the chamber at room temperature. The RGS was then cycled between 0 °C and 70 °C, and the current through the 500-$\Omega$ resistor was measured using a benchtop digital multimeter. The result demonstrated that the current through the resistor varied by only 70 $\mu$A over the entire temperature range, corresponding to a restive measurement uncertainty of $\pm 18$ m$\Omega$, significantly less than the resistance change due to a single line break.

### B. ACS Validation

To properly characterize MMOD based on the reconstructed acoustic signals, it was critical that the acoustic signals were accurately and precisely digitized. The most frequently used assessment of the dynamic performance of an ADC is a spectral analysis based on a sine wave input and fast Fourier transform (FFT) [42]–[45]. To use this technique, however, it is imperative to use a spectrally pure sine wave, where the spurious tones produced by the signal generator fall below the ADC's figure of merits [46]. In addition, the majority of ADC figure of merits are focused on communication systems that have stringent linearity and dynamic range requirements [46]. While important, many of these are unnecessary to characterize for DRAGONS, where our objective is the accurate reconstruction of a specific waveform—the acoustic waveform generated by the impact of a hypervelocity particle. The following ADC characteristics were empirically determined to have the greatest effect on the performance of the ACS system: 1) signal-to-noise ratio referenced to full scale (SNRFS); 2) signal-to-noise and distortion (SINAD); 3) intermodulation distortion (IMD); 4) minimum detectable signal (MDS); 5) spurious free dynamic range (SFDR); and 6) effective number of bits (ENOBs). Results are provided in Table II.

TABLE II
ACS MEASURED PERFORMANCE RESULTS

| Parameter | Value |
|---|---|
| Signal to Noise Ratio Full Scale (SNRFS) | 65.0 dB |
| Signal to Noise and Distortion (SINAD) | 59.8 dB |
| Intermodulation Distortion (IP3) | +22.0 dBFS |
| Minimum Detectable Signal (MDS) | -65.6 dBFS |
| Spurious Free Dynamic Range (SFDR) | 58.4 dB |
| Effective Number of Bits (ENOB) | 9.6 |

*1) Signal-to-Noise Ratio Referenced to Full Scale:* A single-tone test was conducted to determine the SNRFS of the ACS [45]. The $SNRFS$ compares the average power of the desired signal injected at full scale to the total noise power of all spectral components except the first six harmonics and dc (dB). The single-tone $SNRFS$ test consisted of inputting a 20-kHz tone (3.7 Vpp full scale) to one channel of the ACS. The standard ACS measurement of 16 384 samples (used for multiple tests) was then recorded, and the frequency spectrum of the channel was calculated using an FFT.

*2) Signal-to-Noise and Distortion:* SINAD characterizes the degradation of the input signal as a result of both noise and distortion from undesired harmonics [45], [47]. The undesired harmonics consists of the sum of the power in the harmonics within the first Nyquist zone (dB).

*3) Intermodulation Distortion:* IMD, as quantified by the third-order intercept point (IP3) [47], is a result of nonlinearities within the ACS, primarily the amplifier and ADC [47]. The IP3 characterizes the relative strength of internally generated third-order harmonics; these harmonics cause unwanted distortion of the input signal. A two-tone test is the standard technique for measuring IP3; to evaluate the ACS, 20- and 22-kHz signals were generated using an arbitrary waveform generator and input to a single channel of the ACS board. The standard ACS measurement was then recorded, and the spectrum was computed using the FFT.

*4) Minimum Detectable Signal:* The MDS indicates the weakest discernible signal that the system can record [47]. In order to determine the MDS, a signal generator was set to produce a full-scale (0 dBFS) single-tone sine wave. The sine wave was then placed in series with an adjustable attenuator whose output was connected to a single input of the ACS. A standard measurement was recorded, and the output power of the tone was calculated; the attenuation was increased until the output signal power dropped below the noise floor.

*5) Spurious Free Dynamic Range:* SFDR is the strength ratio of the fundamental desired signal to the strongest spurious signal generated by system nonlinearities [45]. SFDR represents the smallest value of an input signal that can be distinguished from a full-scale input signal, and can be calculated after the MDS and IP3 are obtained.

*6) Effective Number of Bits:* The ENOBs quantify the impact of the nonlinearities on the ADC, which is particularly important for time-interleaved sampling structures, because the interleaving process can significantly degrade the ENOB of the system [45]. ENOB is calculated after the $SINAD$ is obtained.

*C. Low-Velocity Impact Error*

Although the spectral analysis metrics provide a general sense of the performance of an ADC and data acquisition system, as well as an easy mechanism to compare data acquisition systems, a sine wave does not illuminate all ADC codes with equal probability [48]. Although a variety of other test methodologies and metrics are presented in the literature [48], of primary concern to DRAGONS is distortion imparted to the actual acoustic waveform as a byproduct of the amplification, filtering, and time-interleaved sampling process. To characterize this distortion, a series of low-velocity impact tests were performed at the United States Naval Academy particle impact facility. The facility consisted of an Airforce R0401 air gun mounted on an aluminum structure firing brass pellets of 0.79 mm (1/32 in) in diameter at a rate of roughly 440 m/s (1450 ft/s). The target was the outer layer of the DRAGONS sensor, that is, four resistive quadrants with four acoustic sensors. The signals from the four acoustic sensors were split simultaneously into two identical amplification circuits. The amplifier outputs were connected to the DRAGONS ACS and the Tektronix MSO4104B mixed-signal oscilloscope and simultaneously captured by both systems. To ensure a proper comparison, both systems were configured to have a sampling frequency of 500 kHz per channel and record 1000 samples per channel. The resistive-grid outputs were split simultaneously into the input of the RGS board and the Fluke 8846A Precision Multimeter to validate the change in resistance compared with the change in digital count output from the RGS board. This enabled the accuracy and precision of the acoustic and resistive subsystems to be evaluated.

To evaluate the accuracy of the ACS, a root mean square error (rmse) comparison of the ACS and MSO4104B digitized waveforms was conducted. A series of ten shots were fired at different locations on the outer layer of the DRAGONS sensor. The average ACS input signal had a maximum voltage of 1 V, which is approximately 11 dB below the full-scale input of the ACS, but well within the dynamic range of the system. For the ACS, the average rmse was 5.3% of the maximum input voltage, with a standard deviation of $\pm 1.2\%$. For the RGS, the rmse was 0.05 $\Omega$ with a standard deviation of $\pm 0.01$ $\Omega$.

IV. HYPERVELOCITY IMPACT PERFORMANCE

The principle goal of the hypervelocity test was to evaluate the prototype DRAGONS ACS configuration characteristics and its ability to measure MMOD impacts. The analysis procedure used was blind, because it could be used to analyze data collected from a space-deployed system. The only *a priori* information used was the locations of the sensors on each layer, the layer separation distance, and the wave speed on the film. Testing was conducted in the University of Kent at Canterbury hypervelocity impact laboratory [49]. The laboratory consisted of a large vacuum chamber and a two-stage light gas gun. The light gas gun had the capacity to fire MMOD-like particles ranging from 25 $\mu$m to 2.5 mm in diameter at velocities as high as 5.9 km/s. In addition, the facility includes a dual-laser system for accurately measuring the projectile and fragment speeds for validation.

*A. Impact Test Setup*

To properly test the DRAGONS system capability to characterize MMOD, a number of variables were configured for each shot in an attempt to replicate the impacts experienced in orbit:
1) angle of impact;
2) material of the particle;
3) size of the particle;
4) speed of the particle.

The DRAGONS frame was attached to a mechanical structure that could rotate on one axis to alter the impact angle, as shown in Fig. 5(a). Common MMOD materials, such as copper, stainless steel (SS), glass, and iron, of different sizes were used to test the system in an environment simulating real-world conditions. Once the projectile and frame orientation were configured for the desired test, the firing speed was selected. The two nominal firing speeds used in our tests were controlled by the type of gas used: 5.2 km/s (hydrogen) and 1.2 km/s (argon). The actual particle velocity was measured

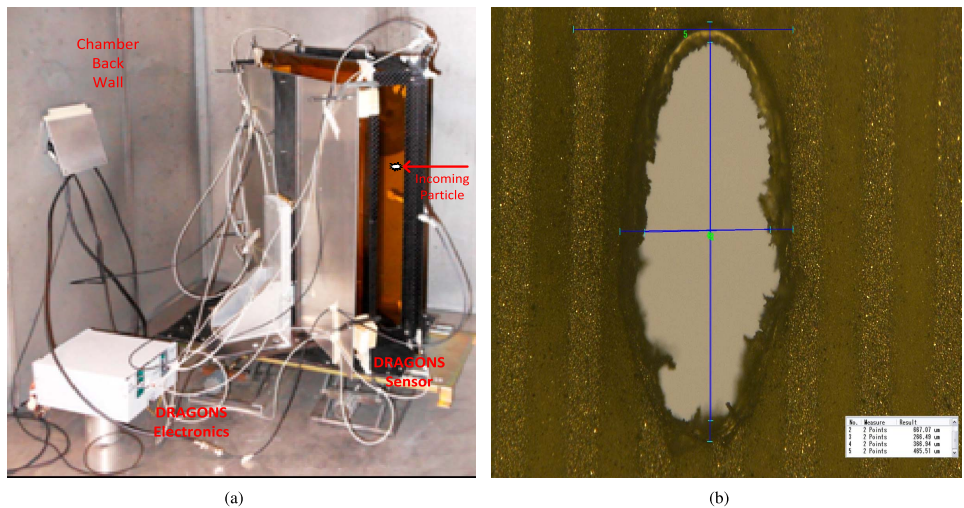

Fig. 5. Image of the setup used for hypervelocity impact testing of the DRAGONS system. (a) DRAGONS instrument installed on a rotating pedestal inside impact test chamber and configured for a 45° off-axis shots. (b) Example of the impact crater caused by a 0.20-mm glass particle fired at $5.2 \pm 0.3$ km/s. The inside dimension of the crater is 670 $\mu$m long and 260 $\mu$m wide.

TABLE III
HYPERVELOCITY SHOT CONFIGURATIONS

| Shot No. | Pellet | Actual AoA (°) | Actual Velocity (km/s) | Measured AoA (°) | Measured Velocity (km/s) |
|---|---|---|---|---|---|
| 1 | 1.0 mm SS | 0 | 5.19 | 5 | 5.4 |
| 2 | 1.0 mm SS | 0 | 5.09 | 6 | 5.2 |
| 3 | 1.0 mm SS | 0 | 5.19 | 4 | 6.8 |
| 4 | 0.8 mm SS | 45 | 1.29 | 56 | 1.1 |
| 5 | 0.8 mm SS | 45 | 5.00 | 51 | 5.5 |
| 6 | 164 $\mu$m Cu | 0 | 5.20 | 3 | 5.2 |
| 7 | 164 $\mu$m Cu | 60 | 1.27 | 55 | 1.3 |
| 8 | 164 $\mu$m Cu | 45 | 1.30 | 35 | 1.8 |
| 9 | 164 $\mu$m Cu | 60 | 5.37 | 59 | 5.8 |
| 10 | 164 $\mu$m Cu | 45 | 5.37 | 46 | 5.1 |
| 11 | 164 $\mu$m Cu | 70 | 5.37 | 63 | 5.3 |
| 12 | 164 $\mu$m Cu | 45 | 1.34 | 48 | 1.3 |
| 13 | 304 $\mu$m Glass | 0 | 5.23 | 11 | 5.4 |
| 14 | 50 $\mu$m Fe | 45 | 1.34 | 42 | 1.3 |

SS - Stainless Steel sphere
Glass - Glass sphere
Cu - Copper sphere
Fe - Iron dust particles

using a dual-laser system installed in the test chamber. A total of 14 shots were completed with the individual shot parameters given in Table III. An example of an impact on the resistive grid is shown in Fig. 5(b).

### B. Postprocessing and Impact Test Results

DRAGONS makes use of a geometric multilateration ToA technique for determining the position of an impact based on the arrival times of acoustic waveforms at the acoustic sensors. Although many multilateration algorithms are commonly available in the literature, all use iterative solving techniques. For example, Cui *et al.* [50] investigate a problem similar to DRAGONS—localizing a gunshot or explosion in a 3-D field. In both cases, the waveforms of interest are fast transients and nondeterministic. They propose two time difference of arrival techniques, a genetic algorithm, and a particle swarm, and are able to demonstrate accurate localization through simulation and experimental results [50]. These iterative techniques can rapidly converge to a solution; however, in processor-limited deployments, they are computationally expensive to implement. For DRAGONS, we derived a simple closed-form analytical expression that makes use of two geometry constraints applicable to our systems: the surface is planar, and the sensors can be located orthogonally.

For all ToA calculations, the wave speed used is 1900 m/s for the kapton-only layer and 2360 m/s for the resistive-grid layer; these values were empirically determined using calibrated exciters attached to each film. Precisely determining the arrival time of each acoustic waveform is crucial to determining the impact location of the particle, from which velocity and AoA will be calculated. The wave speed will, however, change slightly as a result of the temperature and aging of the kapton sheet and acoustic sensors. Although these effects can be measured on the ground prior to launch, we use the previously mentioned piezoelectric disk actuator to provide *in situ* monitoring of wave speed. Essentially, the calibration procedure consists of activating the piezoelectric disk and measuring the relative signal arrival times at each of the acoustic sensor. Because the distance between each sensor and the disk is known, the location of the disk can be calculated using the multilateration procedure (discussed in the following), and the wave speed can be adjusted until the calculated location matches the known location. In addition, the disk actuator can be used to monitor for changes in the sensitivity of the sensors; these changes can be used to adjust the impact-detection threshold.

In principle, signal arrival time is just the first detection of the appearance of the signal—the leading edge. In practice, this is impractical for two reasons: 1) the signals are dispersive and are traveling on lossy thin films and 2) the earliest arrivals are often difficult to clearly separate from noise, particularly with lower amplitude and lower velocity impacts. Use of the leading edge is particularly prone to error if signals are small and simple trigger-threshold type detection is used. Further complicating signal detection arises from DRAGONS' use of very thin materials, the strong influence exerted by the surface on the wave, and the dispersive nature of the waves themselves. This dispersion is more familiar for the case of films under tension (e.g., a kettle drum), where an impact generates resonance (modes) each with unique wave numbers (frequency and wave speed) and damping factor. In our case, the film is not under significant tension, and the modal frequencies are at very low frequencies. These would not appear until long after our signal record is truncated, and would be strongly suppressed by the high-pass filter used in the ACS electronics. While the modal response is not relevant in our signal bandwidth of interest, the wave speed and damping on the film retains its strong frequency dependence. Mathematically, this is expressed as including an imaginary part to the wave speed that can be larger than the real part. The complex wave speed causes a number of interesting but complicating phenomena: 1) the wave envelope travels at a speed (group velocity) that is no longer the same as that of the cyclical troughs and peaks (phase velocity); 2) the higher frequency components of the wave travel faster; and 3) the damping or attenuation increases strongly with frequency. The net result is that the shape of the signal changes considerably as the wave progresses.

More advanced approaches that make use of the waveform shape include determining the envelope and extrapolating back to determine its start and frequency compensating for inhomogeneous materials to reduce distortion. These approaches, however, can be complex and require significant computational resources. The approach used in DRAGONS is a hybrid approach that uses an energy-arrival trigger. The use of the energy trigger partially compensates for the problems associated with lossy inhomogeneous media, where higher frequency components travel faster and are more quickly attenuated, but does not require a significant amount of system complexity or computational power. The key benefit is that simple Euclidean geometry can be used to calculate the location of the particle impact, AoA, and the velocity of a particle impact via

$$L_p = \sqrt{(x_2 - x_1)^2 - (y_2 - y_1)^2 + h^2} \quad (1)$$

$$\theta_x = \arctan\left(\frac{x_2 - x_1}{h}\right) \quad (2)$$

$$\theta_y = \arctan\left(\frac{y_2 - y_1}{h}\right) \quad (3)$$

$$c_p = \frac{L_p}{t_2 - t_1} \quad (4)$$

where $(x_1, y_1)$ is the particle impact location on the top layer, $(x_2, y_2)$ is the particle impact location on the bottom layer, $h$ is the spacing between the layers, $L_p$ is the path length, $\theta_x$ and $\theta_y$ are the AoA on each layer (where 0° is defined as normal to the plane of incidence), and $c_p$ is the particle speed based on the impact time $t_1$ and $t_2$ on each layer.

Once recorded, each acoustic signal was converted into a signal proportional to energy, via

$$e(t) = v^2(t) \quad (5)$$

where $v(t)$ is the recorded voltage signal and $e(t)$ is the resulting energy signal. The energy signal is then integrated to generate a depiction of the evolution of signal energy as a function of time, using

$$e_{\text{sum}}(t) = \int_0^t e(\tau)d\tau. \quad (6)$$

Because the integration process will also incorporate the electrical system background noise level, the background noise contribution is first evaluated from the earliest portion of the data record (prior to the arrival of any acoustic signals)

$$e_n(t) = n^2(t) \quad (7)$$

where $n(t)$ is the recorded background noise signal (taken from the first 10% of samples well before the arrival of the acoustic signal, and $e_n(t)$ is the resulting energy signal). From there, the average noise contribution per unit time is calculated as follows:

$$\overline{N} = \frac{1}{L}\int_0^L e_n(t)dt \quad (8)$$

where $L$ is the length of the noise waveform. This average contribution per interval is then subtracted from the integrated energy signal to create a record that contains only acoustic contributions, as follows:

$$E = e_{\text{sum}}(t) - \overline{N} \quad (9)$$

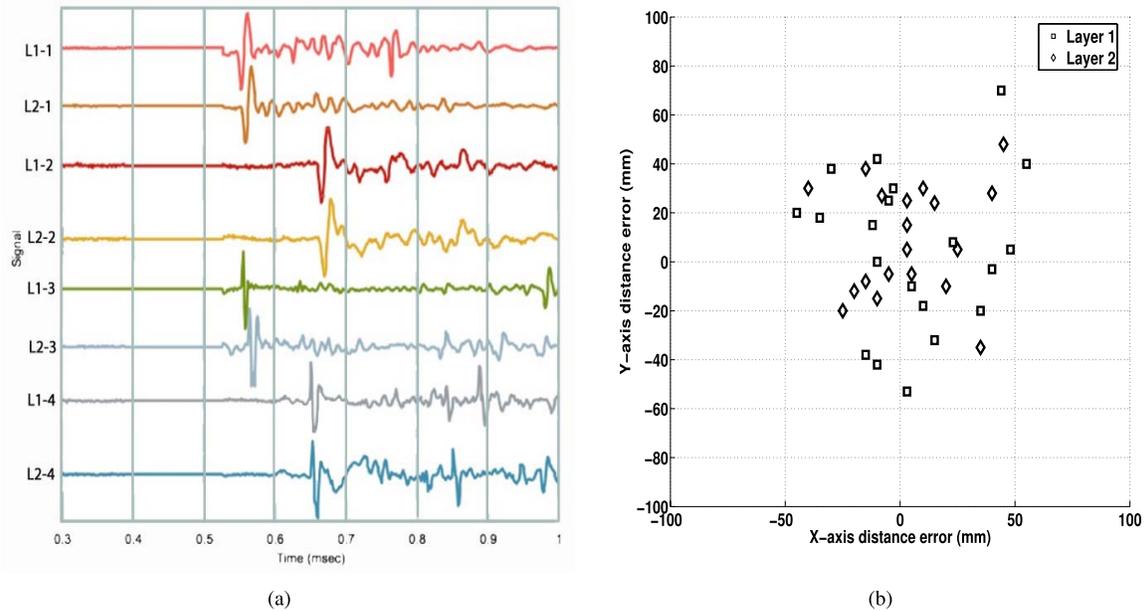

Fig. 6. Results from hypervelocity impact testing of DRAGONS. (a) Example set of acoustic waveforms captured from a hypervelocity impact. L1 represents the top (resistive grid) layer and L2 represents the bottom (kapton-only) layer. Waveforms have been normalized to their maximum value for ease of display. (b) Location errors for the hypervelocity impact tests for both L1 and L2.

where $E$ is referred to here as the integrated acoustic-signal energy function, $E$. A relative energy threshold $T_E$ is then assigned to demarcate the arrival time of the acoustic signal. For most signals, the arrival time is then taken as the time at which the signal reaches $T_E = 15\%$ of its maximum value $E_{MAX} = \max[E]$. This threshold is usually sufficient to avoid early noise contamination from fast high-frequency components.

The hypervelocity impact test data were used to validate the ability of DRAGONS to accurately measure particle velocity and impact location. Fig. 6(a) displays an example set of acoustic signals recorded by the DRAGONS ACS system. Illustrated in Fig. 6(a) are the acoustic signals recorded from the top layer (L1, or the resistive-grid layer) juxtaposed against signals recorded from the bottom layer (L2, or the kapton-only layer). Fig. 6(a) shows that the wave propagation speed in kapton is sufficiently slow to allow for the clear identification of the arrival time of the impact waveform at each acoustic transducer. However, the time it takes a particle to traverse the 15 cm distance between the top and bottom layers is sufficiently small that the accuracy of particle velocity will degrade as the particle velocity increases.

Fig. 6(b) shows the error between the true particle impact location on the top (L1) and bottom (L2) layers and the DRAGONS ACS calculated particle impact location. Fig. 6(b) shows that DRAGONS has a relatively modest error, with a maximum worst case linear position error (in either the $x$- or $y$-direction) of 15% and a more typical linear position error of only 4%. These values translate into an average linear position error of 1.4 and 2.1 cm for the top and bottom layers, and a standard deviation of 2 and 2.9 cm. Finally, we note that the position error appears to be uncorrelated with regard to the impact layer (i.e., knowing the position error on the top layer provides no information about the position error on the bottom layer).

Table III provides the results of the DRAGONS ACS calculated particle velocity and AoA. Table III shows that DRAGONS was able to achieve a mean AoA error of 5° with the standard deviation of 6° for all velocities. Considering only the SS, iron, and glass particles, DRAGONS exhibited a mean velocity measurement error of 0.2 km/s, with a standard deviation of ±0.5 km/s. Of particular importance was the ability to successfully detect and measure extremely small 50-$\mu$m particles. Although the resistive-grid layer provides an independent estimate of particle size, the acoustic system provides the data necessary to determine velocity and AoA, which are the key parameters needed to characterize the particle dynamics. The final hypervelocity shot used 50-$\mu$m iron dust particles, which were clearly detected with an SNR of 24 dB. Interestingly, the 304-$\mu$m glass particles exhibited the weakest SNR of only 13 dB; however, this was still sufficient to ensure the accurate measurement of AoA and velocity.

Although the critical requirement for the DRAGONS ACS system is the detection of signal timing, of secondary importance is determining the relationship between acoustic-signal amplitude and particle size. Accurately determining the particle size via the ACS provides a vital corroboration of the RGS estimate as well as clear evidence of DRAGONS ability to measure the statistics of sub-100-$\mu$m particles that may only cause a single line break in the resistive grid.

Fig. 7(a) shows a comparison between the two speed levels used in the hypervelocity impact testing and the measured peak-to-peak signal amplitude. Fig. 7(b) shows a comparison between the particle diameter and measured signal amplitude, both the "raw" amplitude, as well as $\sqrt{E}$ (integrated acoustic-

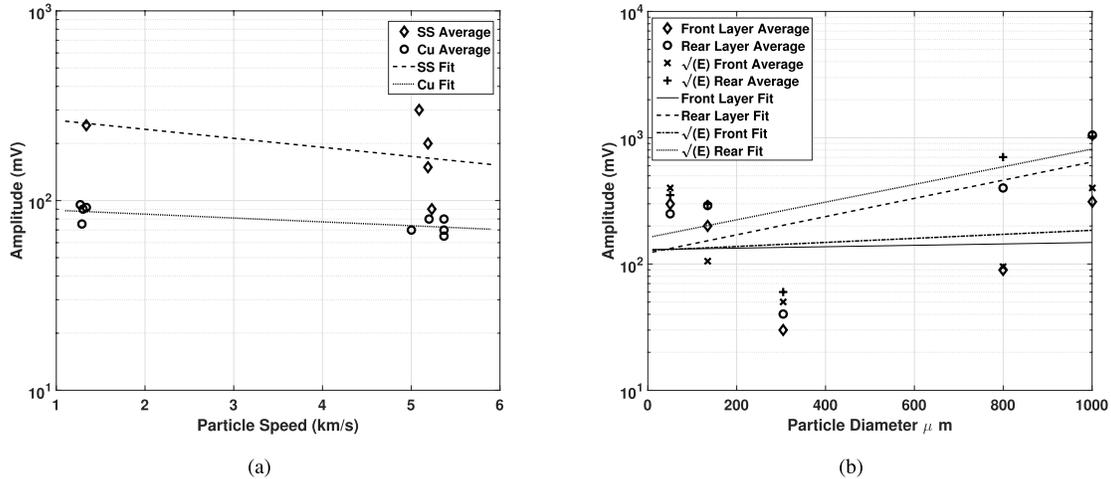

Fig. 7. Measured signal amplitude as a function of particle size and velocity for SS and copper (Cu) particles. Also shown is a linear regression (in a log scale) fit to the measured data. (a) Signal amplitude versus particle speed. (b) Signal amplitude versus particle size.

TABLE IV
SIGNAL CORRELATION WITH PARTICLE TYPE AND VELOCITY

| Parameter | $R^2$ | $F$-Statistic | $p$-value |
|---|---|---|---|
| SS Speed vs. Amplitude | 0.217 | 0.564 | 0.507 |
| Cu Speed vs. Amplitude | 0.532 | 7.94 | 0.026 |
| Particle Diameter vs. Front Amplitude | 0.003 | 0.001 | 0.931 |
| Particle Diameter vs. Rear Amplitude | 0.345 | 1.58 | 0.298 |
| Particle Diameter vs. Front $\sqrt{E}$ | 0.028 | 0.085 | 0.798 |
| Particle Diameter vs. Rear $\sqrt{E}$ | 0.181 | 1.89 | 0.263 |

SS - Stainless Steel sphere
Cu - Copper sphere

signal energy) calculated from (9). For each data set represented in Fig. 7, we applied a linear regression (in a log scale) to determine whether there was a statistically significant relationship between particle size or speed to recorded amplitude or energy, the results of which are presented in Table IV. From Table IV, we observe $R^2$ values ranging from essentially 0–0.5 indicating a weak-to-moderate linear (in log scale) relationship between particle speed or diameter and impact amplitude/energy. However, linear relationship is only statistically significant at the 0.05 level for copper particles. Admittedly, this representation does not present the most unequivocal data set due to the differing conditions for each shot, including different particle materials and various angles of incidence (from 0° to 70°). However, the data and analysis suggest that particle speed does not influence the resulting acoustic-signal level.

Further considering Fig. 7, we can observe complex behaviors that can roughly be summarized as follows.

1) For large particles (0.8- and 1-mm SS), the signal amplitude is weakly linearly related to the particle diameter to within the measurement error and small data set. The 304-$\mu$m glass particle also appears to fit this trend, but because only one test was conducted, it is uncertain whether this is an artifact.

2) The smaller particles (164-$\mu$m copper and 50-$\mu$m iron) exhibited signal amplitudes that were large relative to the glass and SS. Again, the 50-$\mu$m iron represents only one test, so it is uncertain whether this is an artifact.

Although we have limited data with which to draw conclusions, our hypothesis is that fast-moving particles with a diameter larger than the Kapton thickness are essentially unimpeded by the film. Thus, they transfer little energy to the film, which is consistent with the first observation above. As the particle becomes smaller, it encounters progressively increasing resistance when penetrating the film, and thus, it transfers a higher percentage of its energy to the film. As a consequence, the lateral stress and subsequent detection as an acoustic signal would be proportionally increased, which is consistent with the second observation. More testing is, however, necessary in order to firmly establish the relationship between particle size, amplitude, and velocity.

## V. CONCLUSION

Overall, the DRAGONS system demonstrated the ability to detect particle impacts and determine particle location, speed, and AoA with a mean error of 1.4 cm, 0.2 km/s, and 5°. With the increasing need to characterize MMOD, the DRAGONS system could be the a viable solution. The DRAGONS system has been demonstrated to be an accurate and reliable instrument for characterizing the MMOD field. DRAGONS modular, inexpensive, and lightweight design enables it to be installed on most spacecraft in both LEO and GEO. It is one of the first technologies that enable the measurement of both particle velocity and AoA of MMOD. It is vital to improve the understanding of the ever-growing MMOD issue, as the reliance on space technologies continues to expand. The deployment of DRAGONS will be able to fill the important measurement gap that exists in the current debris models.

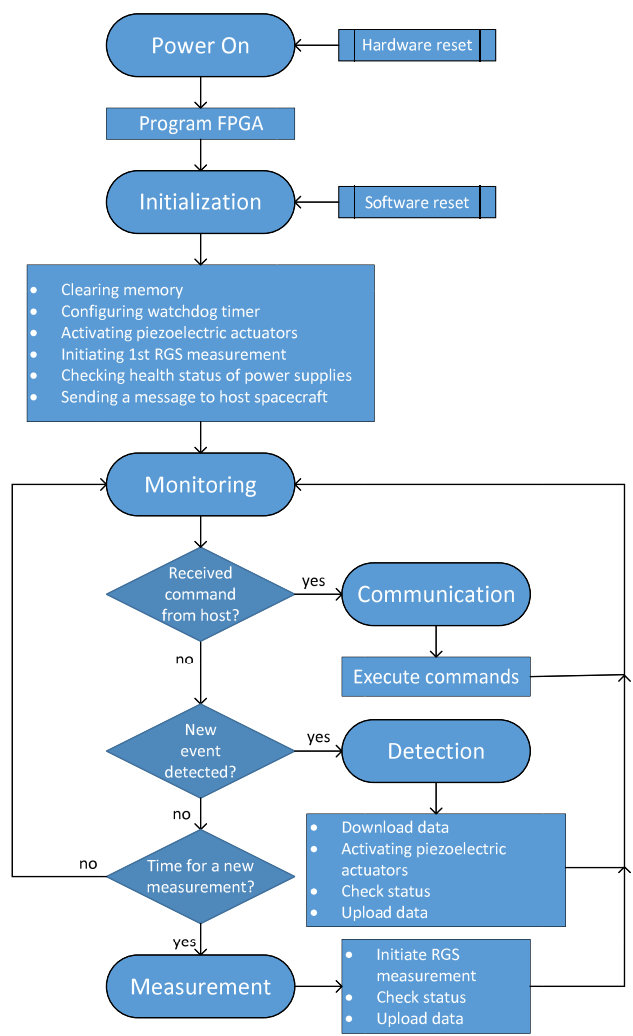

Fig. 8. Flowchart illustrating the operation of the DRAGONS electronics system.

## APPENDIX
## DRAGONS SYSTEM OPERATION

DRAGONS operates in five different states depending on the situation: 1) initialization; 2) monitoring; 3) communication; 4) impact detection; and 5) periodic measurement. These five states are shown in Fig. 8 and discussed in the following. During every action, the date and the time are stored with the data to track the events position in orbit.

1) *Initialization:* During the initialization state, the CDSS programs the FPGAs on all three of the subsystems and clears the FIFOs and SRAM. Afterward, the watchdog timer is initialized, the ACS and RGS are calibrated, and the power supply voltage levels for the entire system are verified.
2) *Monitoring:* In the monitoring state, the CDSS waits for one of three different conditions to occur: communication with the host satellite, impact detection from the ACS, or a periodic health, status, and calibration measurement.
3) *Communication:* If communication is required between DRAGONS and the host satellite, the CDSS will enter the communication state.
4) *Impact Detection:* If the ACS detects a particle impact, CDSS enters the impact-detection state. DRAGONS records the impact data, excites the actuator circuit, records a resistive and temperature measurement, and checks the status of the voltage supplies.
5) *Periodic Measurement:* When the on-board timer alerts the CDSS of a periodic health and calibration measurement, CDSS will enter the periodic measurement state. In this state, a resistive, temperature, and health and status measurement are recorded.

### ACKNOWLEDGMENT

The authors would like to thank V. Pisacane, A. Sadilek, F. Giovane, and J.-C. Liou for sharing their deep technical knowledge throughout this project, as well as J. S. Kang, D. Marasco, H. M. Doyle, and E. Petrosky for their assistance in collecting and analyzing experimental data. In addition, they would also like to thank M. J. Burchell and M. J. Cole at the University of Kent for performing the hypervelocity impact tests.

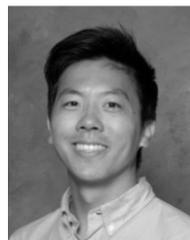

**Michael A. Tsao** received the B.S. degree in electrical engineering from The Pennsylvania State University, State College, PA, USA, in 2012. He is currently pursuing the M.S. degree in computer science with Johns Hopkins University, Baltimore, MD, USA.

He joined the United States Naval Academy, Annapolis, MD, USA, in 2011, as a Research Assistant to Dr. C. Anderson to assist in the development of DRAGONS. This is his second IEEE publication. He is currently a Software Engineer at Northrop Grumman researching computer vision and machine learning.

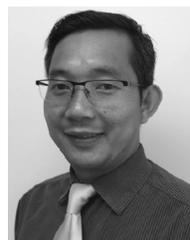

**Hau T. Ngo** (M'08) received the B.S. and M.S. degrees in computer engineering and the Ph.D. degree in electrical and computer engineering from Old Dominion University, Norfolk, VA, USA, in 2001, 2003, and 2007, respectively.

He is currently an Associate Professor with the Electrical and Computer Engineering Department, United States Naval Academy, Annapolis, MD, USA. His current research interests include biometric signal processing, power-delay-area efficient design methodologies for image and video processing systems, real-time signal processing on field programmable gate arrays, hardware–software co-design of high performance embedded systems, and digital forensics.

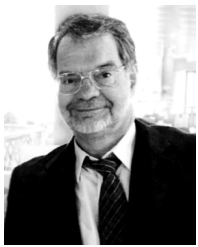

**Robert D. Corsaro** received the Ph.D. degree in physical chemistry from the University of Maryland, College Park, MD, USA, in 1971.

He began his research with the U. S. Naval Research Laboratory, Washington, DC, USA, studying sound propagation in glasses and lossy materials. He retired as the Head of the NRL Wave Effects Section, Herndon, VA, USA, in 2009, but continues to contribute there as a Contractor with Sotera Defense Solutions. He serves as an Independent Contractor and a Consultant for NASA. He holds nine patents, and authored 100 formal publications.

Dr. Corsaro is a fellow of the Acoustical Society of America, where he serves on the Technical Committee for Engineering Acoustics. He is also a member of the American Chemical Society.

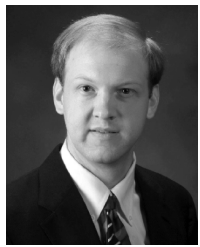

**Christopher R. Anderson** (SM'11) received the B.S., M.S., and Ph.D. degrees in electrical engineering from Virginia Tech, Blacksburg, VA, USA, in 1999, 2002, and 2006, respectively.

He joined the United States Naval Academy (USNA), Annapolis, MD, USA, as an Assistant Professor in 2007. In 2013, he was promoted to Associate Professor of Electrical Engineering. He is currently the Founder and Director of the USNA Wireless Measurements Group, a focused research group that specializes in spectrum, propagation, and field strength measurements in diverse environments and at frequencies ranging from 300 MHz to over 20 GHz. His research has been funded by the National Science Foundation, the Office of Naval Research, NASA, the Defense Spectrum Organization, and the Federal Railroad Administration. He has authored or co-authored over 30 refereed publications. His current research interests include radiowave propagation measurements and modeling, embedded software-defined radios, dynamic spectrum sharing, and ultrawideband communications.

Dr. Anderson is an Editor of the IEEE Transactions on Wireless Communications and a Guest Editor of the IEEE Journal on Selected Areas in Signal Processing of the Special Issue on Non-Cooperative Localization Networks.